\documentclass[a4paper,11pt]{article}
\usepackage{pos}
\usepackage{hyperref}
\usepackage{siunitx}
\usepackage{graphics}
\usepackage[percent]{overpic}
\usepackage{pict2e}
\usepackage{setspace}

\usepackage{wrapfig}

\title{Sensitivity studies for the IceCube-Gen2 radio array}
 \ShortTitle{Sensitivity studies for the IceCube-Gen2 radio array}

\author{The IceCube-Gen2 Collaboration \\{\normalsize \normalfont(a complete list of authors can be found at the end of the proceedings)}}

\emailAdd{steffen.hallmann@desy.de}

\abstract{
The IceCube Neutrino Observatory at the South Pole has measured the diffuse astrophysical neutrino flux up to $\sim$PeV energies and is starting to identify first point source candidates.
The next generation facility, IceCube-Gen2, aims at extending the accessible energy range to EeV in order to measure the continuation of the astrophysical spectrum, to identify neutrino sources, and to search for a cosmogenic neutrino flux. As part of IceCube-Gen2, a radio array is foreseen that is sensitive to detect Askaryan emission of neutrinos beyond $\sim$30 PeV. Surface and deep antenna stations have different benefits in terms of effective area, resolution, and the capability to reject backgrounds from cosmic-ray air showers and may be combined to reach the best sensitivity. The optimal detector configuration is still to be identified.

This contribution presents the full-array simulation efforts for a combination of deep and surface antennas, and compares different design options with respect to their sensitivity to fulfill the science goals of IceCube-Gen2.\\

\vspace{4mm}
{\bfseries Corresponding authors:}
Steffen Hallmann$^{a,*}$, Brian Clark$^{b}$, Christian Glaser$^{c}$, Daniel Smith$^{d}$\\[4mm]
{$^{a}$ \itshape Deutsches Elektronen-Synchrotron (DESY),
Platanenallee 6, 15738 Zeuthen, Germany}\\
{$^{b}$ \itshape Michigan State University, Department of Physics and Astronomy, 
567 Wilson Rd, East Lansing, MI 48824, USA}\\
{$^{c}$ \itshape Uppsala Universitet, Departement of Physics and Astronomy, Ångströmlaboratoriet, 
Lägerhyddsvägen 1, 751 20 Uppsala, Sweden}\\
{$^{d}$ \itshape University of Chicago, Department of Physics, 
5720 S Ellis Ave, Chicago, IL 60637, USA}
\\[4mm]
$^*$ Presenter
}

\FullConference{37$^{\rm{th}}$ International Cosmic Ray Conference (ICRC 2021)\\
		July 12th -- 23rd, 2021\\
		Online -- Berlin, Germany}

\begin{document}
\maketitle

\section{A radio array as part of IceCube-Gen2}
The IceCube Neutrino Observatory has convincingly measured the presence of a diffuse astrophysical neutrino flux \cite{icecube_diffuse}, and has identified first likely candidates for point source emission through multimessenger observations \cite{icecube_txs, icecube_tde}. 
In order to reveal the contributors to the diffuse astrophysical neutrino flux and to push the energy frontier up to EeV energies, a next generation extension, IceCube-Gen2 \cite{gen2_window_extreme_universe}, is planned.
Scaling the current optical Cherenkov technique to reach the latter goal is complicated by the relatively short  $\mathcal{O}(\SI{100}{m}$) attenuation length of optical light in the ice.
An array of radio antennas is therefore planned as part of IceCube-Gen2 to detect the Askaryan emission of neutrino-induced particle showers in the frequency range between \SI{100}{MHz} to \SI{1}{GHz}. The radio signal (similar to the Cherenkov light) has a cone-shaped emission profile. Thanks to the larger attenuation length of radio emission -- $\mathcal{O}$(1\,km) -- a sparsely instrumented detector with $\mathcal{O}$(500) stations covering an area of $\sim\SI{500}{km^2}$ on the ice can therefore obtain sufficient sensitivity to astrophysical and cosmogenic neutrinos despite the steeply falling neutrino flux with an energy threshold at $\sim\SI{30}{PeV}$.

IceCube-Gen2 Radio benefits from the experience gained from all previous ground- and balloon-based \cite{ANITA} radio neutrino experiments.
In the benchmark scenario presented in these proceedings, the detector combines high-gain log-periodic dipole antennas (LPDAs) buried close to the snow surface, as used in ARIANNA \cite{ARIANNA_detector}, with up to 200\,m deep vertically (Vpol) and horizontally (Hpol) polarized dipole antennas, as used in ARA \cite{ARA_detector}, similar to the station design of
the Radio Neutrino Observatory in Greenland (RNO-G) \cite{rnog}. 

In this work, we present the simulation efforts for a benchmark array in \autoref{sec:simulated_array}. In \autoref{sec:sensitivities}, it is shown that the simulated array layout is sensitive to detect cosmogenic neutrinos and probe astrophysical neutrino emission at the highest energies. The performance of the instrument in terms of coincident triggers, analysis efficiency, background rejection and signal reconstruction is discussed in \autoref{sec:eff_area_reco}. \autoref{sec:additional_studies} concludes with an outlook how to further optimize the design.


\section{Simulated radio array}\label{sec:simulated_array}
A complete detector array with a total of 313 stations has been simulated as a benchmark radio scenario for IceCube-Gen2.
The software packages \texttt{NuRadioMC} \cite{NuRadioMC} and \texttt{NuRadioReco} \cite{NuRadioReco} have been used for the generation and propagation of the Askaryan radio signal, and for the simulation of the detector response and event reconstruction, respectively. The simulation takes into account additional radio signals that are generated by muons and taus that were created by charge current interactions of the respective neutrino \cite{secondary_interactions}. 

\begin{figure}
    \centering
    \begin{minipage}{.5\textwidth}
        \includegraphics[width=\columnwidth]{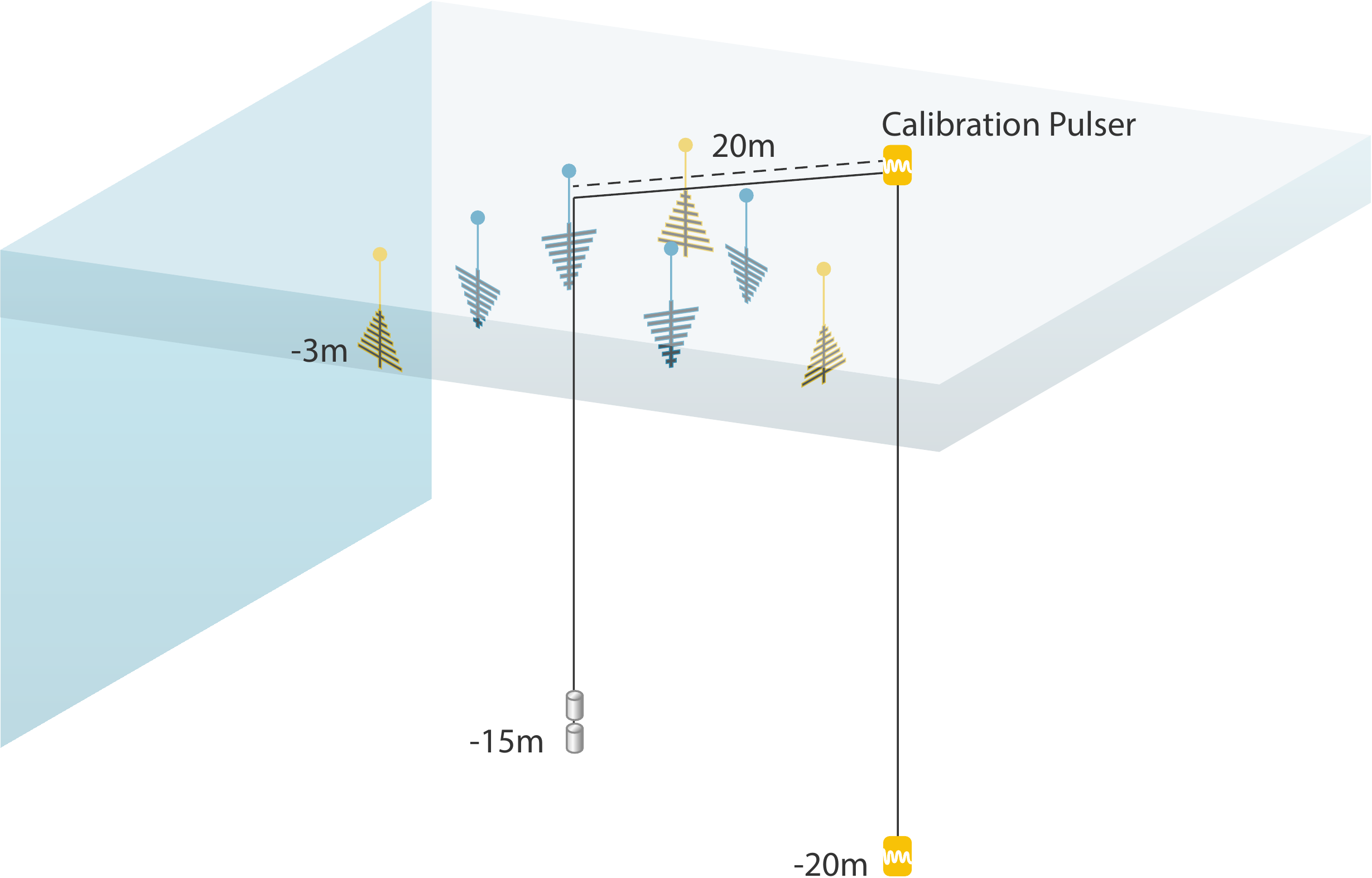}
        \caption{\textbf{(top)} Layout of a shallow detector station. It consists of 4 downward facing LPDAs and a vertically polarized dipole antenna at \SI{15}{m}. These are complemented by 3 upward facing LPDAs to measure (and veto) cosmic rays as well as anthropogenic noise.\vspace{.5\baselineskip}
        \newline
        \textbf{(right)} Layout of a hybrid station featuring a shallow component (as in the shallow detector station) and a deep component. The latter consists of a 200\,m deep phased array of vertically polarized dipole antennas for triggering and additional antennas for reconstruction with vertical and horizontal polarization response.}
        \label{fig:simulated_station}
    \end{minipage}    
    \hfill
    \begin{minipage}{.43\textwidth}
        \centering
        \begin{overpic}[width=\columnwidth
        ]{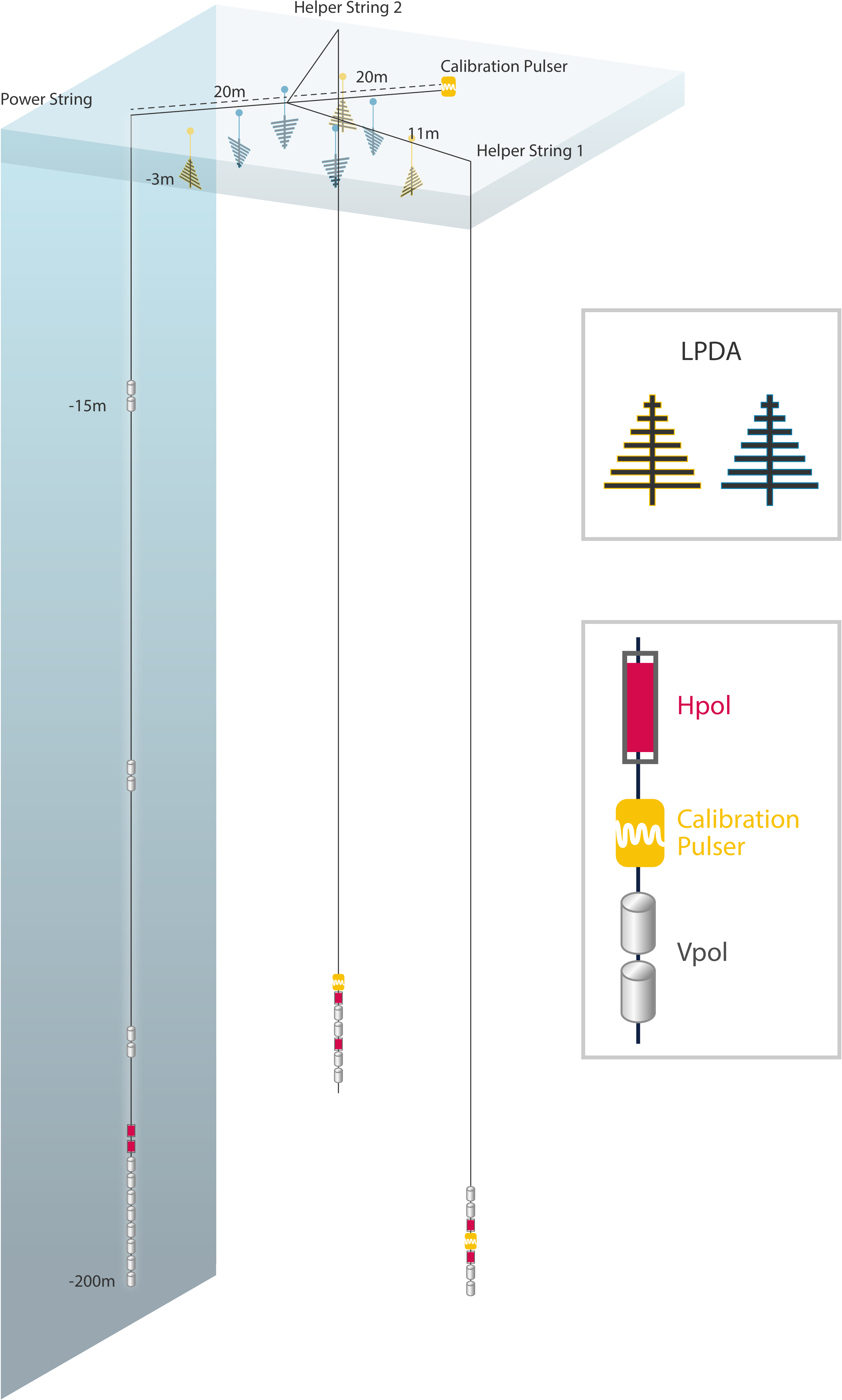}
        \end{overpic}

    \end{minipage}
\end{figure}

The benchmark array extends over an area of $\sim\SI{500}{km^2}$. 
It consists of a square array of 144 (12$\times$12) hybrid stations horizontally spaced by 2\,km. Hybrid stations are equipped with both a deep and a shallow component as shown in Fig.~\ref{fig:simulated_station}. In addition, 169 (13$\times$13) stations with only a shallow component are placed between and around the hybrid stations.
The hybrid stations consist of 24 detection channels, while the shallow stations have only 8 channels.

The radio array is designed such that each detector station operates autonomously. Neutrino events are triggered and can be reconstructed with a single station. This allows to instrument huge volumes cost efficiently with a sparse array of radio detector stations.

In the shallow component, downward and upward facing LPDAs are complemented by a shallow Vpol antenna at \SI{15}{m} depth. In-ice showers are triggered in the shallow component by a bandwidth optimized coincidence trigger on the downward facing LPDA antennas \cite{Bandwidthpaper}. An additional trigger runs on the upward facing LPDAs for cosmic-ray detection. 
The depth of the dipole is chosen so that the majority of events will have a visible \emph{DnR} signature where two time-delayed pulses are observable in the dipole antenna. The first pulse originates from a direct signal trajectory to the antenna whereas the second pulse is reflected off the surface at total internal reflection or refracted for most geometries. This provides a unique signature of an Askaryan signal.

The deep component uses an interferometric trigger on four Vpol antennas that are placed vertically above each other. The signals of the four dipoles are digitized in real-time and multiple beams are formed to cover the elevation angle range expected for neutrino signals from roughly \SI{-60}{\degree} to \SI{60}{\degree}. This beamforming increases the signal-to-noise ratio by up to a factor of two (=$\sqrt{4})$ compared to a single antenna. This concept was pioneered by ARA and allows to lower the detection threshold of the instrument \cite{ara_phased_array_trigger,ICRC2021_Kaeli_phased_array}. Additional Vpol and Hpol antennas are placed above the phased array and on two additional strings to aid event reconstruction following the RNO-G design. 

\section{Sensitivity to the diffuse neutrino flux and to point sources}\label{sec:sensitivities}
%

\begin{figure}
    \centering
    \begin{minipage}{.49\columnwidth}
    \includegraphics[width=\columnwidth]{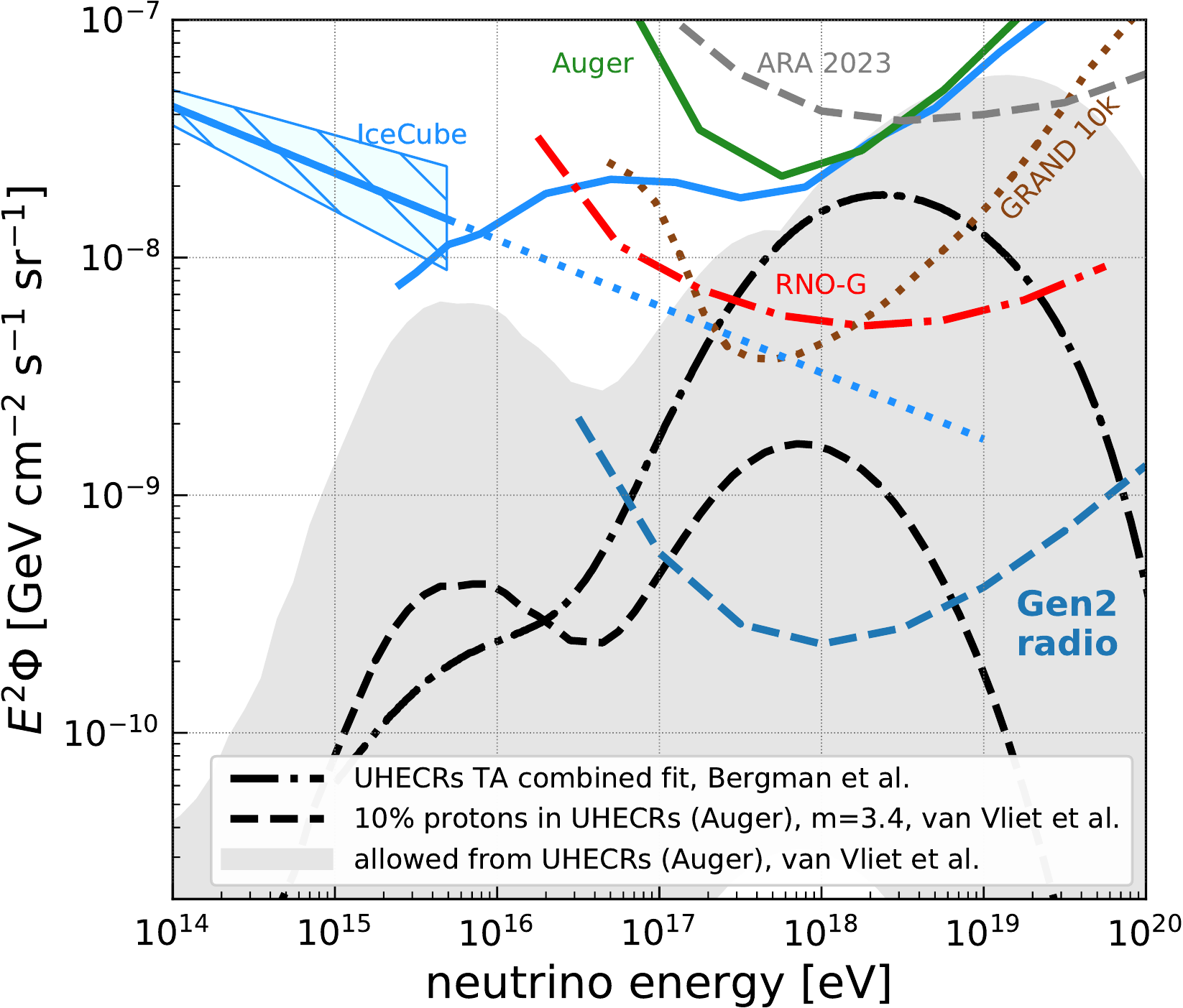}
    \caption{The ten year expected differential 90\% CL sensitivity at trigger level for zero background of the simulated radio array to a diffuse neutrino flux is shown as dashed line. Solid lines show the astrophysical neutrino flux measured by IceCube \cite{icecube_spectrum} and experimental upper limits at higher energies. The expected sensitivities of ARA (for 2023), of RNO-G currently under construction and the proposed GRAND10k array (both for ten years) are also shown, as well as different predictions of the GZK neutrino flux based on UHECR data.
    }
    \label{fig:diffuse_sensitivity}    
    \end{minipage}
    \hfill
    \begin{minipage}{.47\columnwidth}
    \includegraphics[width=\columnwidth]{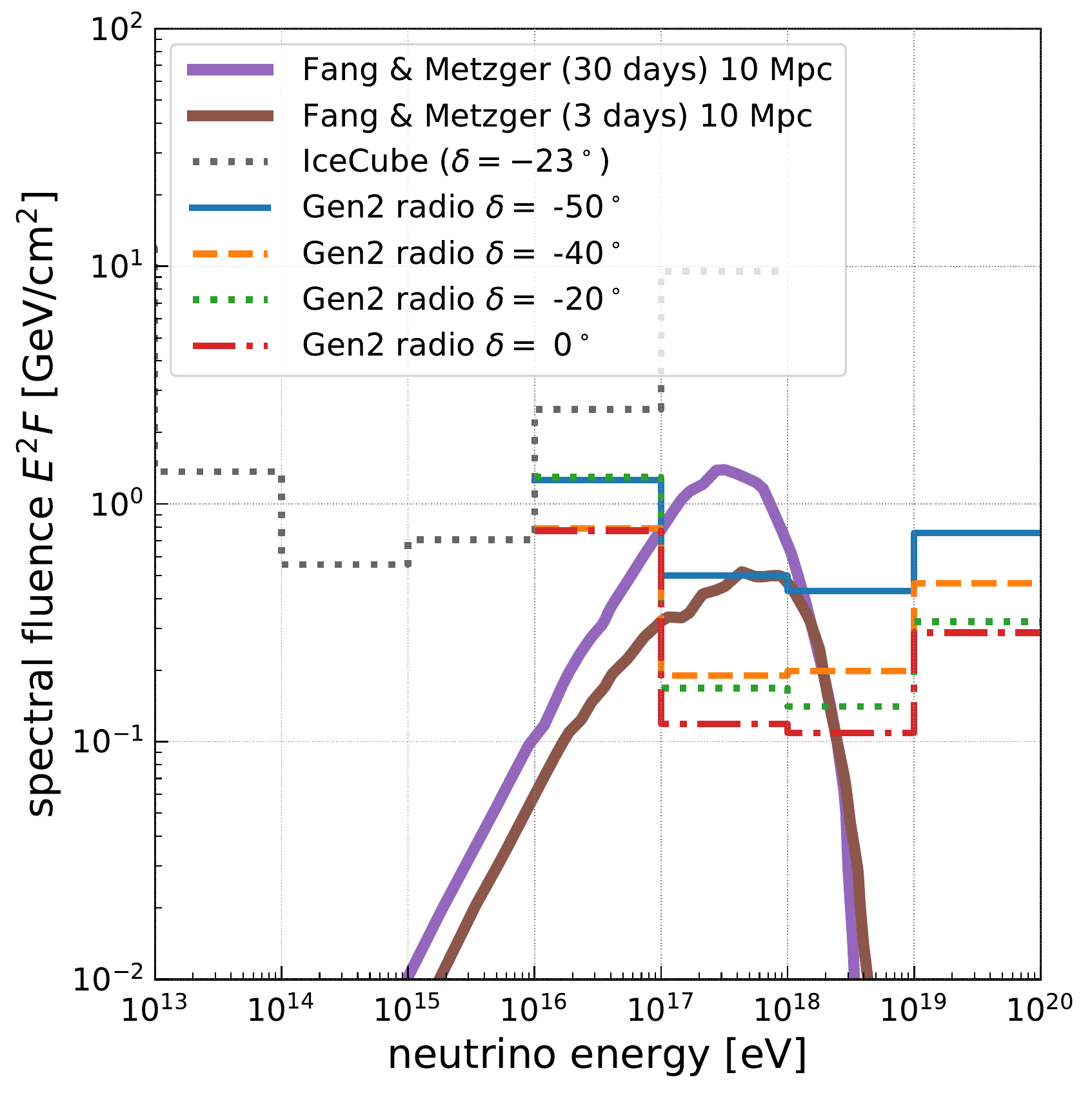}
    \caption{90\% CL fluence sensitivity for the IceCube-Gen2 Radio array for transient point sources located at different positions on the sky. Fluence predictions of neutron star -- neutron star mergers as detected by gravitational wave observations \cite{fang_metzger} are added for comparison.}
    \label{fig:ps_sensitivity}    
    \end{minipage}
\end{figure}

To get an overview of the expected sensitivity, we show the trigger level sensitivity on a diffuse all-flavor neutrino flux for a zero background hypothesis that can be achieved with the simulated benchmark detector for 10 years of uptime in \autoref{fig:diffuse_sensitivity}. The sensitivity is shown differentially for decade wide bins in neutrino energy. We discuss background and the rejection strategies in \autoref{sec:analysis_efficiency_background}.

The predicted fluxes of GZK neutrinos show large variations. The best fit to cosmic-ray data of the Telescope Array will result in close to 240 detected neutrinos within ten years of operation \cite{TA}, whereas a fit of the same model with a 10\% proton contribution to data of the Pierre Auger Observatory yields only 20 detected neutrinos \cite{vanvliet_neutrinos}. 
Also, the number of detected diffuse astrophysical neutrinos strongly depends on the continuation of the neutrino spectrum measured by IceCube. For an unbroken astrophysical neutrino spectrum that follows $E^{-2.28}$, as the one shown in \autoref{fig:diffuse_sensitivity}, the radio detector of Gen2 will measure 74 neutrinos in ten years where most detected neutrinos will have energies between \SI{e17}{eV} and \SI{e18}{eV}.


The sensitivity to transient point sources is shown for different source positions in \autoref{fig:ps_sensitivity}. The instrument is mostly sensitive between $\delta\approx-40^\circ$ and $\delta=0$ as the Earth is opaque to neutrinos at ultra-high energies (UHEs), i.e. energies $\gtrsim\SI{10}{PeV}$. Due to the location at the South Pole, the same part of the sky is observed continuously. The large instantaneous sensitivity paired with an almost background free measurement of neutrinos at UHEs will allow searches for transient events. The detector will participate in the growing network of multi-messenger observations. As one promising example, a model of neutrino production in a neutron star -- neutron star merger (NSNS) is shown in \autoref{fig:ps_sensitivity}, where the rate of detected NSNS will increase significantly with the new gravitational wave detectors coming online in the next years.  

The detector will be designed to identify neutrinos in real time and to alert other telescopes for follow up observations with short latency. This will enable multi-messenger astronomy with neutrinos at ultra-high energies and could potentially lead to a direct discovery of a source of ultra-high energy cosmic rays (UHECRs).


\section{Performance of the simulated benchmark array}\label{sec:eff_area_reco}
The benchmark design is evaluated in more detail in terms of coincident detection of events, analysis efficiency, background rejection capabilities and reconstruction performance.

\begin{wrapfigure}{r}{8cm}
    \centering
    \includegraphics[width=7.9cm]{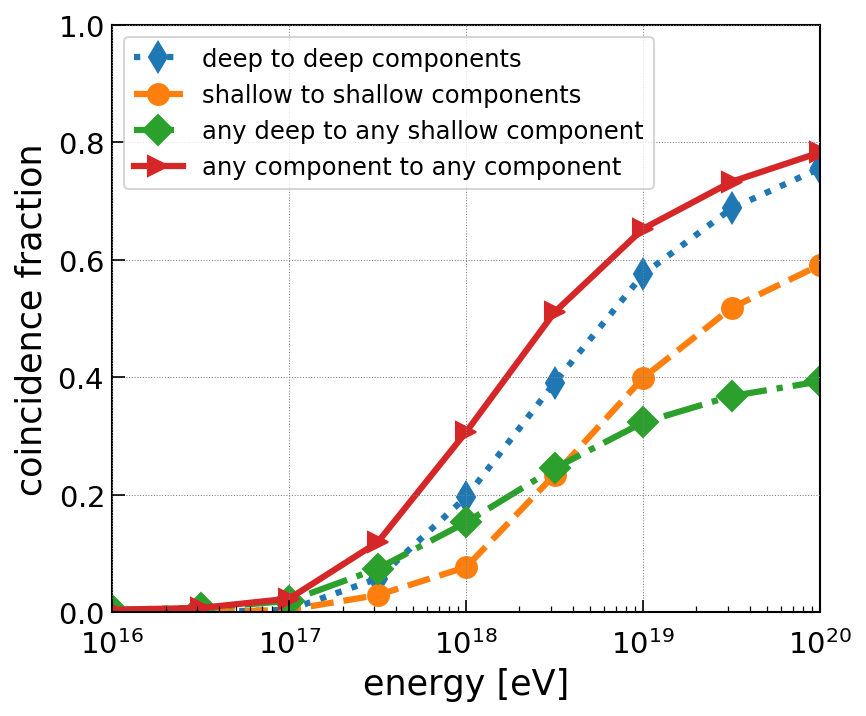}
    \caption{Fraction of events with multiple triggers as a function of neutrino energy. Coincident triggers on more than one deep ('deep to deep') and shallow ('shallow to shallow') component are seen at least in two different stations. This is not necessarily the case for events triggering at least one deep and one shallow component ('any deep to any shallow') or any two components ('any component to any component').}
    \label{fig:coincidences}
\end{wrapfigure}



\subsection{Coincident triggers}

As the radio array of Gen2 is a discovery instrument and given that the flux predictions of UHE neutrinos have large uncertainties, the detector design puts emphasis on redundancy. The shallow and deep components will allow to measure neutrinos with largely complementary systematic uncertainties (e.g. different regions of the ice) and the spacing between the stations is adjusted such that a subset of neutrinos will be detected in multiple detector stations simultaneously. The corresponding fractions of coincident triggers between deep and shallow components and on multiple stations are given in \autoref{fig:coincidences} for different neutrino energies. This golden event sample will allow a detailed characterization of the radio signatures of UHE neutrinos, and provide an improved event reconstruction.  

At the same time, an increasing overlap between stations will decrease the overall sensitivity of the detector. The spacing of \SI{2}{km} between hybrid stations is a  compromise between maximizing sensitivity and building redundancy into the detector. At the benchmark energy of \SI{e18}{eV} the fraction of events with coincident triggers in the detector is 30\%. 

The inter-station coincidences provide two unique opportunities for event reconstruction: First, since different stations will see a different part of the Cherenkov cone, they can provide a sample of events with improved angular resolution. Second, multiple showers beyond PeV are expected not only for $\nu_\tau$ charged current interactions (the characteristic 'double-bang' signature) but also from catastrophic losses of $\mu$ and $\tau$ leptons along their trajectory \cite{secondary_interactions}. Hence, the detection of multiple spatially separated showers can be used for flavor identification and to improve event reconstruction. Work to quantify the improvements enabled by these coincident events is ongoing.

\subsection{Analysis efficiency and background rejection}\label{sec:analysis_efficiency_background}
The detector is designed to effectively suppress backgrounds. The most common background is thermal noise fluctuations that fulfill the trigger condition since trigger thresholds are set close to the thermal noise floor at $\sim 2 \sigma$ for a \SI{100}{Hz} trigger rate. Other common types of noise sources are anthropogenic and wind-related background. These types of noise are readily rejected with high efficiency as demonstrated in ARA and ARIANNA.  

Estimates based on ARA data for the deep component with a phased array trigger \cite{ICRC2021_Kaeli_phased_array} show that thermal backgrounds can be suppressed efficiently while retaining 
68\% and 83\% at 
$\SI{e17}{eV}$  and $\SI{e18}{eV}$ of the triggered neutrino signals. 
These numbers are likely to improve by $\mathcal{O}(20\%)$ assuming additional gains in the analysis and help of the shallow component. 
Similarly, ARIANNA showed that -- in a shallow station that comprises only four downward facing LPDAs -- thermal, wind-related and anthropogenic backgrounds can be rejected while retaining $\sim79\%$ of all triggered neutrino signals \cite{ARIANNA_4p5y_analysis}. With the additional dipole antenna that provides the unique DnR signature and the upward facing LDPA antennas, the number is expected to improve to close to 100\% with respect to trigger level. 

A physics background of concern originates from cosmic rays, where three types of backgrounds need to be suppressed: the radio signal generated in the air primarily through geomagnetic emission, the radio signal generated by the Askaryan effect from an incompletely developed air showers continuing in ice, and the one from catastrophic losses of high energy atmospheric muons that produce $>$PeV showers in ice. 
The dense surface instrumentation with upward facing LPDA antennas allows for cosmic-ray tagging to suppress this background. For the simulated station spacing, the array reaches full efficiency around \SI{1}{EeV} for inclined air showers ($>70^\circ$ zenith angle), which is the relevant zenith range for the muon background. With roughly \SI{500}{km^2}, the radio array of IceCube-Gen2 will be a relatively large air shower detector, about a fifth of the size of the Pierre Auger Observatory. 
Absolute in-ice background estimates induced by atmospheric muons depend strongly on the cosmic-ray composition and the interaction model used. Especially the prompt muon production is largely unconstrained at these high energies and theoretical models have significant uncertainties. Using current models (the GSF cosmic-ray flux \cite{GSF} and Sibyll 2.3c \cite{sibyll23c}), we expect 0.4 events from high-energy muons passing through the ice per year for the complete array. 
A thorough estimation of potential background and array optimizations to mitigate them efficiently are pursued with high priority at the moment. 

While the cosmic-ray induced radio signals constitute a background for neutrino detection, they also act as a good calibration source and provide unique science cases. Because the radio emission from air showers is well understood, and can be measured efficiently with the upward facing antennas, it can be used as a continuous in-situ calibration source with properties similar to the Askaryan signals expected from neutrino interactions. 
Furthermore, the combined measurement of the air shower via its in-air emission and the in-ice signal from a catastrophic energy loss of an atmospheric muon provides the unique opportunity to measure the muon production at UHE. 

\subsection{Direction and energy resolution}
The determination of the neutrino energy and direction requires the reconstruction of several observables due to the characteristics of the Askaryan emission. The neutrino direction requires a reconstruction of the signal arrival direction, the viewing angle, i.e., the angle under which the shower is observed, as well as the polarization of the signal. A poor polarization measurement will result in a banana-shaped uncertainty region on the sky.  To determine the neutrino energy, also the distance to the neutrino vertex is needed as well as an absolute calibration of the detector to recover the signal that arrived at the antenna. In addition, the ice needs to be understood to correct for attenuation and bending of signal trajectories. 

Studies from ARIANNA show that the shallow detector can achieve an average angular error of around three degrees for all triggered events \cite{ICRC2021_ARIANNA_direction}. With additional quality cuts, the resolution can be further improved. The two pairs of orthogonal LPDAs allow reconstruction of the signal polarization in a way that allows most systematic uncertainties in the antenna response to cancel out. The broadband response enables the reconstruction of the viewing angle, and the signal direction is determined via triangulation. Signal direction and polarization reconstruction was confirmed in in-situ measurements \cite{ARIANNA_polarization}. The \SI{15}{m} deep dipole will measure a direct signal and a second signal pulse from a reflection off the surface for almost all events (\emph{DnR} signature). This aids the reconstruction of the viewing angle by mapping out the Cherenkov cone. The time difference between these \emph{DnR} pulses is also an estimator of the distance to the neutrino vertex and enables resolving the shower energy with an accuracy well below the intrinsic fluctuations from the unknown\footnote{Unless for $\nu_e$ charged current, where both the hadronic and electromagnetic showers are observed.} inelasticity of the interaction which amounts to a factor of two \cite{ARIANNA_energy_resolution}. 

Studies from ARA and RNO-G inform the expected performance of the deep detector. 
The vertex position can be determined from the signal arrival times of the 3D array of antennas and the frequency content provides information about the viewing angle. ARA has demonstrated that through triangulation the direction to the vertex can be determined to within a degree
~\cite{Abdul:2017luv}. 
Studies for RNO-G also indicate that after quality cuts that assure a good enough signal-to-noise ratio also in the antennas surrounding the phased array trigger, the shower energy can be determined to a precision better than the intrinsic uncertainty from inelasticity fluctuations \cite{Welling_energy_reco} (which dominate the energy reconstruction). 
The reconstruction of the neutrino direction is more challenging in the deep component because of the smaller sensitivity of the Hpol antennas compared to the Vpol (dipole) antennas that are used for triggering. For a subset of events where also the Hpol antennas measure a large enough signal strength (or provide a significant enough null measurement), the neutrino direction can be reconstructed with a few degrees uncertainty as well \cite{Plaisier_direction_reco}; the impact of systematic uncertainties in the different antenna responses is still under study.


\section{Conclusion and further layout optimization}\label{sec:additional_studies}
We have presented simulations for the radio array of IceCube-Gen2. It searches for radio emission from in-ice particle showers initiated by neutrino interactions. The radio technique allows a cost efficient instrumentation of huge volumes with a sparse array of autonomous detector stations. The detector will provide unprecedented sensitivity to UHE neutrinos with a peak sensitivity between \SI{e17}{eV} and \SI{e19}{eV} and will probe even pessimistic models of high-energy neutrino production. The large instantaneous sensitivity will allow searches for transient events and participate in multi-messenger campaigns which could lead to the direct discovery of sources of UHECRs.

As a discovery instrument for UHE neutrinos, emphasis is put on redundancy and background rejection. In particular, the deep and shallow detector components will allow triggering, identification and reconstruction of UHE neutrinos with complementary systematics; the station spacing is adjusted to have a subset of golden events observed in multiple stations; and the denser surface instrumentation yields an efficient cosmic-ray veto.

The final configuration of the IceCube-Gen2 Radio array is still being optimized.
The use of an additional phased array trigger in the shallow component, and a deep phased array trigger made from 8 instead of 4 Vpols has been studied, providing the potential of further lowering the trigger threshold. With current methods, the increase in sensitivity at analysis level is comparatively small, so further work is needed. This scheme also requires additional channels and is therefore only favorable in case the power consumption is not significantly increased. 

Although the current benchmark design already fulfills the science requirements, further optimization is foreseen. This includes the relative positioning of antennas within a station to further optimize the event reconstruction capabilities, potentially increasing the trigger level sensitivity by increasing the phased array channels from four to eight antennas, and further optimizations of the station spacing with emphasis on cosmic-ray vetoing. 
These studies will guide the path towards a final configuration for the IceCube-Gen2 Radio detector.

\bibliographystyle{ICRC}
\setlength{\bibsep}{2.5pt}
\bibliography{literature.bib}

\clearpage
\section*{Full Author List: IceCube-Gen2 Collaboration}




\scriptsize
\noindent
R. Abbasi$^{17}$,
M. Ackermann$^{71}$,
J. Adams$^{22}$,
J. A. Aguilar$^{12}$,
M. Ahlers$^{26}$,
M. Ahrens$^{60}$,
C. Alispach$^{32}$,
P. Allison$^{24,\: 25}$,
A. A. Alves Jr.$^{35}$,
N. M. Amin$^{50}$
% ...
\\
\\
% ...
$^{12}$ Universit{\'e} Libre de Bruxelles, Science Faculty CP230, B-1050 Brussels, Belgium \\
% ...
$^{17}$ Department of Physics, Loyola University Chicago, Chicago, IL 60660, USA \\
% ...
$^{22}$ Dept. of Physics and Astronomy, University of Canterbury, Private Bag 4800, Christchurch, New Zealand \\
% ...
$^{24}$ Dept. of Astronomy, Ohio State University, Columbus, OH 43210, USA \\
$^{25}$ Dept. of Physics and Center for Cosmology and Astro-Particle Physics, Ohio State University, Columbus, OH 43210, USA \\
$^{26}$ Niels Bohr Institute, University of Copenhagen, DK-2100 Copenhagen, Denmark \\
% ...
$^{32}$ D{\'e}partement de physique nucl{\'e}aire et corpusculaire, Universit{\'e} de Gen{\`e}ve, CH-1211 Gen{\`e}ve, Switzerland \\
% ...
$^{35}$ Karlsruhe Institute of Technology, Institute for Astroparticle Physics, D-76021 Karlsruhe, Germany  \\
% ...
$^{50}$ Bartol Research Institute and Dept. of Physics and Astronomy, University of Delaware, Newark, DE 19716, USA \\
% ...
$^{60}$ Oskar Klein Centre and Dept. of Physics, Stockholm University, SE-10691 Stockholm, Sweden \\
% ...
$^{71}$ DESY, D-15738 Zeuthen, Germany \\
% ...

\subsection*{Acknowledgements}

\noindent
USA {\textendash} U.S. National Science Foundation-Office of Polar Programs,
U.S. National Science Foundation-Physics Division,
U.S. National Science Foundation-EPSCoR,
Wisconsin Alumni Research Foundation,
Center for High Throughput Computing (CHTC) at the University of Wisconsin{\textendash}Madison,
Open Science Grid (OSG),
Extreme Science and Engineering Discovery Environment (XSEDE),
Frontera computing project at the Texas Advanced Computing Center,
U.S. Department of Energy-National Energy Research Scientific Computing Center,
Particle astrophysics research computing center at the University of Maryland,
Institute for Cyber-Enabled Research at Michigan State University,
and Astroparticle physics computational facility at Marquette University;
Belgium {\textendash} Funds for Scientific Research (FRS-FNRS and FWO),
FWO Odysseus and Big Science programmes,
and Belgian Federal Science Policy Office (Belspo);
Germany {\textendash} Bundesministerium f{\"u}r Bildung und Forschung (BMBF),
Deutsche Forschungsgemeinschaft (DFG),
Helmholtz Alliance for Astroparticle Physics (HAP),
Initiative and Networking Fund of the Helmholtz Association,
Deutsches Elektronen Synchrotron (DESY),
and High Performance Computing cluster of the RWTH Aachen;
Sweden {\textendash} Swedish Research Council,
Swedish Polar Research Secretariat,
Swedish National Infrastructure for Computing (SNIC),
and Knut and Alice Wallenberg Foundation;
Australia {\textendash} Australian Research Council;
Canada {\textendash} Natural Sciences and Engineering Research Council of Canada,
Calcul Qu{\'e}bec, Compute Ontario, Canada Foundation for Innovation, WestGrid, and Compute Canada;
Denmark {\textendash} Villum Fonden and Carlsberg Foundation;
New Zealand {\textendash} Marsden Fund;
Japan {\textendash} Japan Society for Promotion of Science (JSPS)
and Institute for Global Prominent Research (IGPR) of Chiba University;
Korea {\textendash} National Research Foundation of Korea (NRF);
Switzerland {\textendash} Swiss National Science Foundation (SNSF);
United Kingdom {\textendash} Department of Physics, University of Oxford.

\end{document}